\documentclass[review]{elsarticle}
\usepackage{color}
\usepackage{hyperref}
\usepackage{amsmath}
\usepackage{amsthm}
\usepackage{multirow}
\usepackage{graphicx}
\usepackage{subfigure}
\usepackage[]{caption2}
\usepackage{booktabs}
\usepackage{geometry}
\usepackage{xcolor}
\usepackage[normalem]{ulem}
\usepackage{gensymb}
\usepackage{makecell}
\usepackage{booktabs}
\newcommand\hl{\bgroup\markoverwith {\textcolor{yellow}{\rule[-.5ex]{2pt}{2.5ex}}}\ULon}
\begin{document}

\renewcommand{\figurename}{Fig.}
\renewcommand{\captionlabeldelim}{.}
\biboptions{numbers,sort&compress}
\newcommand{\upcite}[1]{\textsuperscript{\textsuperscript{\cite{#1}}}}

\begin{frontmatter}

\title{Rheological synergistic thermal conductivity of CMC-based Fe$_{3}$O$_{4}$ and Al$_{2}$O$_{3}$ nanofluids in shear flow fields}
\author[1,2]{Shengna Liu}
\author[2]{Liancun Zheng\corref{corresponding author}}
\ead{liancunzheng@ustb.edu.cn}

\cortext[corresponding author]{Corresponding author.Tel.: +86(10)62332002(Liancun Zheng)}

\address[1]{School of Energy and Environmental Engineering, University of Science and Technology Beijing, Beijing 10083, China}
\address[2]{School of Mathematics and Physics, University of Science and Technology Beijing, Beijing 100083, China}
\begin{abstract}
{\color{black}}In this paper, considering the variation of the viscous dissipative heat in the transfer direction, a new theoretical formula for thermal conductivity measurement was proposed based on the energy equation of the rotational Couette flow field. This theoretical formula shows that thermal conductivity and rheology have a synergistic effect. Based on this theoretical formula, the rheological synergistic thermal conductivity of CMC-based Fe$_{3}$O$_{4}$ nanofluids and Al$_{2}$O$_{3}$ nanofluids was experimentally investigated. The experimental results show that the thermal conductivity rises with the increase of shear rate and volume fraction, moreover, volume fraction and shear rate have mutually reinforcing effects on thermal conductivity enhancement. Non-Newtonian effects of rheology and heat transfer reduce with shear rate and increase with volume fraction, with consistent synergistic effects. According to the experimental data, the expressions of the thermal conductivity and dynamic viscosity of these two nanofluids as functions of shear rate and volume fraction were presented.  \\
\textbf{Key words:} Nanofluid, viscous dissipative heat, rheological synergy, thermal conductivity.\\
\textbf{} \\
\textbf{ }

\end{abstract}

\end{frontmatter}

\section{Introduction}
\label{Introduction}
Energy transportation is a key issue for most industries, such as: petroleum, chemical reactors, nuclear, solar, power plants, microelectronics, etc \cite{keblinski2008thermal,nikkam2014experimental,ozerincc2010enhanced,mahian2013dispersion,shukla2012thermal,bashirnezhad2015comprehensive}. With the trend of miniaturization of industrial equipment, researchers have been exploring the development of new heat exchange fluids to achieve efficient ultra-high performance cooling and remove excess heat \cite{keblinski2008thermal,nikkam2014experimental,patel2008model}. Choi et al. \cite{choi1995enhancing} first named fluids containing nanoparticles as nanofluids. Since Choi et al. proposed the concept of nanofluids, extensive research has been conducted on the thermal properties of nanofluids. Compared with the base fluid, nanofluids can greatly enhance the thermal physical properties, such as thermal conductivity, thermal diffusion coefficient, viscosity and convective heat transfer coefficient, etc. The research on the enhancement mechanism of thermal conductivity of nanofluids is mostly based on theory and experiment. Al$_{2}$O$_{3}$ nanofluids and Fe$_{3}$O$_{4}$ nanofluids are often used as heat transfer fluids or other industrial applications, therefore their thermal conductivities have been extensively studied. Hrishikesh et al. \cite{patel2010experimental} used transient hot-wire equipment and temperature oscillation equipment to experimentally investigate the effects of temperature, nanoparticle size, and volume fraction on the thermal conductivity of water-based Al$_{2}$O$_{3}$ nanofluids. Sundar et al. \cite{sundar2013experimental,sundar2014thermal,sundar2013thermal} conducted experimental studies on the thermal conductivity of ethylene glycol and water mixture-based Al$_{2}$O$_{3}$ nanofluids and Fe$_{3}$O$_{4}$ nanofluids at different temperatures and volume fractions. Pang et al. \cite{pang2012thermal} measured the thermal conductivity of methanol-based Al$_{2}$O$_{3}$ nanofluids using transient hot-wire method. Tavman et al. \cite{tavman2008experimental} examined the influence of particle concentration and temperature on the thermal conductivity of water-based Al$_{2}$O$_{3}$ nanofluids using the 3$\omega$ approach. Afrand et al. \cite{afrand2016experimental} experimentally measured the thermal conductivity of water-based Fe$_{3}$O$_{4}$ nanofluids at different temperatures and volume fractions by employing a KD2 Pro thermal properties analyser. Sundar et al. \cite{sundar2013investigation} studied the effective thermal conductivity and viscosity of water-based Fe$_{3}$O$_{4}$ nanofluids by experimental measurements and theoretical analysis.
\par The above studies on the thermal conductivity of Fe$_{3}$O$_{4}$ nanofluids and Al$_{2}$O$_{3}$ nanofluids were all at static conditions, and the base fluids of the nanofluids were mostly Newtonian fluids. If the base fluid of the nanofluid is a non-Newtonian fluid, it will exhibit non-Newtonian properties. Since carboxymethyl cellulose (CMC) solution has the properties of bonding, colloidal protection, emulsification, suspension and easy flow, and is widely used in petroleum, food, medicine, textile and papermaking industries. CMC-based non-Newtonian nanofluids have attracted the attention of many scholars and a great deal of research has been conducted \cite{zainith2021experimental,saqib2018natural,bazdidi2020evaluation,al2019numerical}. For non-Newtonian fluids, their rheological properties change with the shear rate, implying that the molecular structure of the fluid and the intermolecular interaction forces vary with the shear rate, which inevitably leads to a modification of the thermal conductivity, that is, there is a synergistic coupling between rheology and heat transfer. Based on this, it is necessary to study the thermal conductivity of non-Newtonian fluids in a shear flow state. Previously, most of the research on the thermal conductivity of non-Newtonian fluids in a shear flow field employed the coaxial cylindrical method \cite{broniarz2009thermal,shin2000thermal,lee1997shear,lee1998shear,kostic1999investigation,ikhu2009effect,sun2013shear}. However, none of them took into account the viscous dissipative heat generated by the shear flow of non-Newtonian fluids, resulting in low thermal conductivity measurements. Liu et al. \cite{liu2021rheological} studied the thermal conductivity of HEC-based silica nanofluids under shear flow by using a laboratory-made coaxial cylindrical thermal conductivity measuring device, and for the first time considered the influence of viscous dissipative heat on the thermal conductivity. But Liu et al. regarded the viscous dissipative heat transferred in the annular gap of the coaxial cylinders as the viscous dissipative heat generated in the entire annular gap, ignoring the variation of the viscous dissipative heat in the transfer direction. In fact, the viscous dissipative heat through the annular gap is different at each cross section.
\par In this paper, considering the variation of viscous dissipative heat in the transfer direction, a new theoretical formulation for thermal conductivity measurements is established based on the energy equation of the rotational Couette flow field. According to the theoretical formulation, the rheological synergistic thermal conductivity of CMC-based Fe$_{3}$O$_{4}$ nanofluids and Al$_{2}$O$_{3}$ nanofluids was experimentally investigated using the laboratory-built coaxial cylinder thermal conductivity measurement device and the commercial Anton Paar MCR 302 rheometer.

\section{Preparation of Fe$_{3}$O$_{4}$ and Al$_{2}$O$_{3}$ nanofluids based on 2$wt\%$ carboxymethyl cellulose (CMC) solution}
We prepared Fe$_{3}$O$_{4}$ nanofluids with volume fraction $\phi$=1.195\%, 2.361\%, 3.500\%, 4.614\% and Al$_{2}$O$_{3}$ nanofluids with volume fraction $\phi$=1.313\%, 2.597\%, 3.846\%, 5.063\%, respectively. The base solution of nanofluids is 2wt\% carboxymethyl cellulose aqueous solution. These nanofluids were prepared using a 25wt\% nano Fe$_{3}$O$_{4}$ dispersion purchased from Shanghai McLean Biochemical Technology Co., Ltd. and a 20wt\% nano Al$_{2}$O$_{3}$ dispersion purchased from Shanghai Aladdin Biochemical Technology Co., Ltd. The method of nanofluids preparation is given in the following by taking Fe$_{3}$O$_{4}$ nanofluid with a volume fraction of 1.195\% as an example:
\par The density of Fe$_{3}$O$_{4}$ is 5.18 g/ml and water is 1 g/ml. Assuming that 20 ml nano Fe$_{3}$O$_{4}$ dispersion contains $x$ ml water and $y$ ml Fe$_{3}$O$_{4}$, then x+y=20. Moreover, since the mass fraction of nano Fe$_{3}$O$_{4}$ dispersion is 25\%, therefore $\frac{{5.18{\text{y}}}}{{x + 5.18y}} \times 100\%  = 25\%$. By solving the two equations we get $x \approx 18.791,\ y \approx 1.209$. Thus, a total of $\frac{{1.209}}{{0.01195}} - 1.209 \approx 100$ ml of aqueous solution and carboxymethyl cellulose powder were required for the preparation of Fe$_{3}$O$_{4}$ nanofluid with a volume fraction of 1.195\%. The density of carboxymethyl cellulose is 1.05 g/ml, Assuming there is $z$ g carboxymethyl cellulose powder and $f$ g water in the base liquid, then $\frac{{\text{z}}}{{1.05}} + f = 100$. Because the mass fraction of carboxymethyl cellulose in the base solution is 2\%, it can be obtained that $\frac{z}{{f + z}} \times 100\%  = 2\%$. So $z \approx 2$, $f \approx 98.095$. As 20 ml Fe$_{3}$O$_{4}$ dispersion already contains 18.791 ml water, 98.095-18.791 = 79.3 ml additional water is needed. Add 2 g carboxymethyl cellulose powder and 79.3 ml water to the beaker, stir thoroughly until the carboxymethyl cellulose powder is completely dissolved. Then 20 ml of the purchased nano Fe$_{3}$O$_{4}$ dispersion was added to the prepared CMC solution, stirred thoroughly and sonicated with an ultrasonic cell grinder for several minutes until the nanoparticles were evenly dispersed in the solution. In this way, Fe$_{3}$O$_{4}$ nanofluid with a volume fraction of 1.195\% was prepared using 2wt\% CMC solution as the base solution. The preparation methods of the other nanofluids are similar.
\par Figs. 1(a) and 1(b) show the particle size distribution of Fe$_{3}$O$_{4}$ nanofluids and Al$_{2}$O$_{3}$ nanofluids measured by Zeta potential analyzer, respectively. From the figures, it can be seen that in the three measurements, the particle size distribution of Fe$_{3}$O$_{4}$ nanofluids is around 194 nm, and the particle size distribution of Al$_{2}$O$_{3}$ nanofluids is around 413.6 nm, with a uniform distribution. Through observation, it was found that all nanofluid samples have no obvious deposition within one week.
\begin{figure}[!htb]
 \begin{minipage}[t]{0.5\linewidth}
  \centering
  \centerline{\includegraphics[width=6.5cm, height=4.2cm]{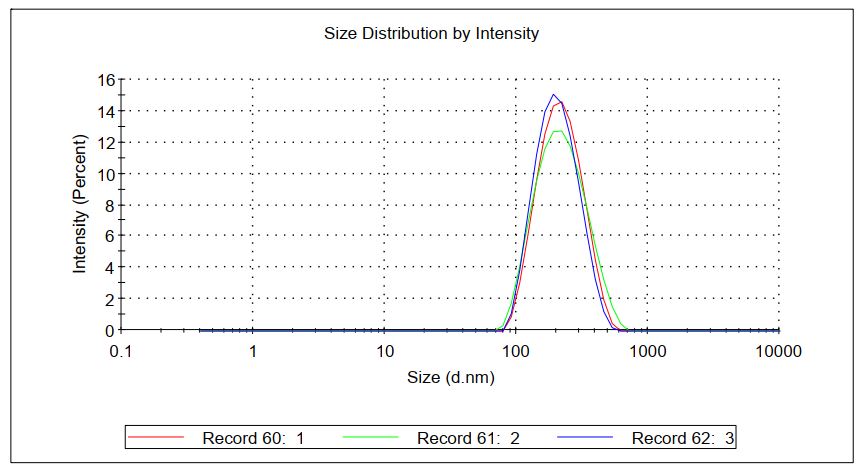}}
 \centerline{(a)}
 \end{minipage}
 \begin{minipage}[t]{0.5\linewidth}
  \centering
  \centerline{\includegraphics[width=6.5cm, height=4.2cm]{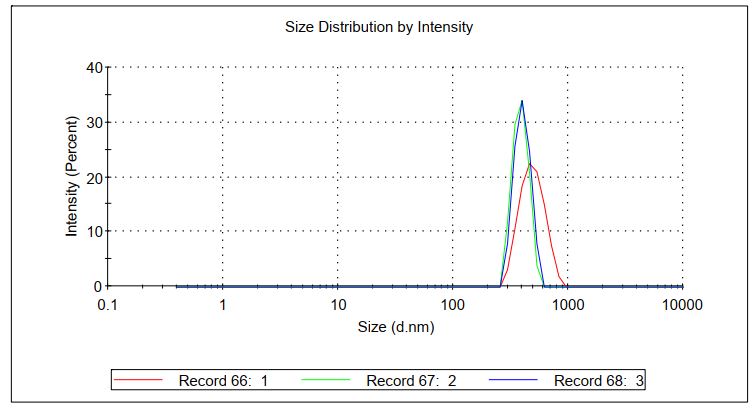}}
  \centerline{(b)}
 \end{minipage}
\caption{Particle size distributions of Fe$_{3}$O$_{4}$ nanofluids (a) and Al$_{2}$O$_{3}$ nanofluids (b).}
\label{fig14}
\end{figure}
\section{{\color{black} Theory of Thermal Conductivity Measurement}}
Based on the rotational Couette flow field established in reference \cite{liu2021rheological}, considering the variation of viscous heat dissipation transmitted radially in the annular gap, the theoretical formula for measuring thermal conductivity considering viscous heat dissipation is derived. The steady-state energy equation of the rotating Couette flow field in a cylindrical coordinate system is:
\begin{equation}
\label{1}
\frac{1}{r}\frac{d}{{dr}}\left( {kr\frac{{dT}}{{dr}}} \right) + {\tau _{r\theta }}\dot \gamma = 0.
\end{equation}
The shear rate in cylindrical coordinate system is represented by
\begin{equation}
\label{2}
\dot \gamma  = \frac{{d{u_\theta }}}{{dr}} - \frac{{{u_\theta }}}{r},
\end{equation}
where $r$ represents the radius, m; $k$ is the thermal conductivity, W/(m$\cdot$K); $T$ is the temperature of the fluid, K; $u_\theta$ is the tangential velocity, m/s; and $\tau _{r\theta }$ is the shear stress, Pa.
\par Ignoring the variation of shear rate $\dot \gamma$ in the radial direction within the annular gap, the shear rate can be approximated as \cite{liu2023heat}:
\begin{equation}
\label{3}
\dot \gamma= \frac{{\pi n{R_{o}}{R_{i}}}}{{15\left( {R_{o}^2 - R_{i}^2} \right)}}
\end{equation}
where $R_{o}$ is the inner radius of the outer cylinder, m; $R_{i}$ is the outer radius of the inner cylinder, m; $n$ is
the rotational speed of the outer cylinder, r/min.
\par The first term on the left side of energy equation (\ref{1}) is the heat conduction term and the second term is the viscous dissipation term. The viscous dissipative heat cannot be ignored when high viscosity fluids flow at high shear rates. ${\tau _{r\theta }}\dot \gamma$ is the viscous dissipative heat generated by shear flow of fluids in the $\theta$ direction per unit volume per unit time. All the viscous dissipative heat generated by the fluid from the outer wall of the inner cylinder to the radius $r$ per unit time is as follows:
\begin{equation}
\label{5}
{Q_2} = \iiint\limits_{V_{1}} {{\tau _{r\theta }}\dot \gamma dV} = {\tau _{r\theta }}\dot \gamma \Delta V_{1} = {\tau _{r\theta }}\dot \gamma \left( {\pi r^2h - \pi R_i^2h} \right) = \pi h{\tau _{r\theta }}\dot \gamma \left( {r^2 - R_i^2} \right),
\end{equation}
where $h$ is the effective heating length of the inner cylinder, m; $\Delta V_{1}=\pi r^2h - \pi R_i^2h$ is the volume of the annular gap from the outer wall of the inner cylinder to the radius $r$. Assuming that the heat generated by heating the inner cylinder per unit time, i.e., the heating power of the inner cylinder is $Q_{1}$. When reaching thermal equilibrium, the heat flow rate $Q$ through the cylindrical side at the radius $r$ position in the annular gap is the heat generated by heating the inner cylinder per unit time, plus the viscous dissipative heat generated from the outer wall of the inner cylinder to the radius $r$ position, that is:
\begin{equation}
\label{6}
Q={Q_1}+{Q_2}.
\end{equation}
Eq. (\ref{5}) shows that the viscous dissipative heat through the side of the cylinder at the position of radius $r$ varies with the radius $r$. Because both the viscous dissipative heat and the heat generated by the inner cylinder are transferred through heat conduction, if the viscous dissipated heat is regarded as the heat flow generated by the inner cylinder, when the thermal equilibrium is reached, there is:
\begin{equation}
\label{7}
\frac{1}{r}\frac{d}{{dr}}\left( {kr\frac{{dT}}{{dr}}} \right) = 0.
\end{equation}
By integrating Eq. (\ref{7}), we get:
\begin{equation}
\label{8}
kr\frac{{dT}}{{dr}} = C.
\end{equation}
Where
\begin{equation}
\label{9}
C =  - \frac{Q}{{2\pi h}} =  - \frac{{{Q_1} + {Q_2}}}{{2\pi h}} =  - \frac{{{Q_1}}}{{2\pi h}} - \frac{{{\tau _{r\theta }}\dot \gamma \left( {{r^2} - R_i^2} \right)}}{2}
\end{equation}
From the Eqs. (\ref{8}) and (\ref{9}), we can get:
\begin{equation}
\label{10}
k\frac{{dT}}{{dr}} =  - \frac{{{Q_1}}}{{2\pi rh}} - \frac{{{\tau _{r\theta }}\dot \gamma \left( {{r^2} - R_i^2} \right)}}{{2r}}
\end{equation}
Integrating Eq. (\ref{10}) from $R_i$ to $R_o$ yields:
\begin{equation}
\label{11}
\int_{{T_i}}^{{T_o}} {kdT = \int_{{R_i}}^{{R_o}} {\left( { - \frac{{{Q_1}}}{{2\pi rh}} - \frac{{{\tau _{r\theta }}\dot \gamma }}{2}r + \frac{{{\tau _{r\theta }}\dot \gamma R_i^2}}{{2r}}} \right)} } dr,
\end{equation}
where $T_{o}$ is the temperature of the inner wall of the outer cylinder, K; $T_{i}$ is the temperature of the outer wall of the inner cylinder, K. By solving Eq. (\ref{11}), the theoretical formula for measuring thermal conductivity considering viscous heat dissipation is obtained as:
\begin{equation}
\label{12}
k = \frac{{{Q_1}\ln \left( {{{{R_o}} \mathord{\left/
 {\vphantom {{{R_o}} {{R_i}}}} \right.
 \kern-\nulldelimiterspace} {{R_i}}}} \right)}}{{2\pi h\left( {{T_i} - {T_o}} \right)}} + \frac{{{\tau _{r\theta }}\dot \gamma \left[ {\left( {R_o^2 - R_i^2} \right) - 2R_i^2\ln \left( {{{{R_o}} \mathord{\left/
 {\vphantom {{{R_o}} {{R_i}}}} \right.
 \kern-\nulldelimiterspace} {{R_i}}}} \right)} \right]}}{{4\left( {{T_i} - {T_o}} \right)}}
\end{equation}
\par It can be clearly seen from formula (\ref{12}) that the thermal conductivity and rheological properties are synergistically coupled, and the thermal conductivity is related to the rheological constitutive equation, i.e. the expression of shear stress. This synergistic correlation is more significant for non-Newtonian fluids whose molecular structure and intermolecular interaction forces change with the shear rate. The traditional formula of linear Fourier heat conduction law is no longer applicable to the heat transfer calculation of non-Newtonian fluids in shear flow state. The rheological properties of non-Newtonian fluids are affected by shear rate, temperature and volume fraction, etc. If the rheological constitutive relationship ${\tau _{r\theta }} = \eta \left( {\dot \gamma ,T,\phi ...} \right)\dot\gamma$ of fluid is known, that is, the function expression of dynamic viscosity $\eta$ with respect to shear rate $\dot \gamma$, temperature $T$, volume fraction $\phi$ and other variables is obtained, the theoretical formula (\ref{12}) for thermal conductivity measurement considering viscous heat dissipation can be written as follows:
\begin{equation}
\label{13}
k = \frac{{{Q_1}\ln \left( {{{{R_o}} \mathord{\left/
 {\vphantom {{{R_o}} {{R_i}}}} \right.
 \kern-\nulldelimiterspace} {{R_i}}}} \right)}}{{2\pi h\left( {{T_i} - {T_o}} \right)}} + \frac{{{\eta \left( {\dot \gamma ,T,\phi ...} \right)}\dot \gamma^{2} \left[ {\left( {R_o^2 - R_i^2} \right) - 2R_i^2\ln \left( {{{{R_o}} \mathord{\left/
 {\vphantom {{{R_o}} {{R_i}}}} \right.
 \kern-\nulldelimiterspace} {{R_i}}}} \right)} \right]}}{{4\left( {{T_i} - {T_o}} \right)}}
\end{equation}
Formula (\ref{13}) reveals that shearing rate, temperature, volume fraction and other factors affect the contribution of viscous dissipation to thermal conductivity. And the viscous dissipation effect of Newton fluid is proportional to the square of the shear rate.
\section{{\color{black}Thermal conductivity measurement steps}}
A laboratory-made coaxial cylinder thermal conductivity measuring device was used to measure the thermal conductivity, and this measurement equipment was described in detail in reference \cite{liu2021rheological}. The measurement steps are as follows:\\
(i) Inject the liquid to be tested into the annular gap between the inner and outer cylinders of the thermal conductivity measurement equipment, with the liquid height consistent with the inner cylinder height.\\
(ii) Set the temperature of the circulating water bath, and then wait for the temperature to reach the set value, and the temperature of the outer walls of the inner and outer cylinders to be consistent.\\
(iii) Set the rotation speed of the outer cylinder $n$ and turn on the motor switch. The shear rate $\dot\gamma$ corresponding to $n$ can be calculated by Eq. (\ref{3}).\\
(iv) Set the heating power of the inner cylinder and turn on the DC power supply (the heating power should not be too large, and ensure that the temperature difference between the outer walls of the inner and outer cylinders does not exceed 10$^{\circ}$C when the temperatures reach stable values).\\
(v) When the system reaches thermal equilibrium, record the values of several thermal resistance thermometers, and calculate the average value to obtain the temperature $T_{i}$ of the outer wall of the inner cylinder and the temperature of the outer wall of the outer cylinder. The temperature $T_{o}$ of the inner wall of the outer cylinder is approximated by the temperature of the outer wall of the outer cylinder.\\
(vi) The average values of $T_{i}$ and $T_{o}$ are used to approximate the fluid temperature. The shear stress at this temperature and shear rate $\dot\gamma$ is measured by Anton Paar MCR 302 rheometer.\\
(vii) substitute the experimental data obtained in the first six steps into Eq. (\ref{12}) to calculate the thermal conductivity considering viscous heat dissipation.\\
(viii) Repeat from (iii) if you need to measure thermal conductivity at different shear rates, and repeat from (i) if you need to measure thermal conductivity of different liquids.\\

\section{{\color{black} Error analysis of thermal conductivity measurement}}
\par Before conducting error analysis, we first made an improvement to the laboratory-made coaxial cylinder thermal conductivity measurement equipment. Wrap an insulation layer on the outer side and top of the circulating water bath barrel of the coaxial cylinder thermal conductivity measurement equipment to reduce the interference of room temperature on the temperature of the circulating water bath. Therefore, in all the measurement results in this paper, the temperature difference between the measured outer cylinder temperature and the set circulating water bath temperature didn't exceed 0.2$^{\circ}$C.
\par If there is free convection in the annular gap, it will cause measurement error. For low Rayleigh number (Ra$<$1000), free convection does not affect heat transfer \cite{sun2013shear,eckertf1961natural}. The expression of Rayleigh number is as follows:
\begin{equation}
\label{14}
Ra = \frac{{g{\beta }\Delta T{{\left( {{R_{o}} - {R_{i}}} \right)}^3}}}{{\alpha \nu }},
\end{equation}
where\ $g$ is the acceleration of gravity, m/s$^2$; $\beta$ is the coefficient of volume expansion, 1/K; $\Delta T$ is the temperature difference between the outer wall of the inner cylinder and the inner wall of the outer cylinder, K; $\alpha$ is the thermal diffusion coefficient, m$^2$/s; $\nu$ is kinematic viscosity (ratio of dynamic viscosity to density), m$^2$/s. For all CMC-based Fe$_3$O$_4$ and Al$_2$O$_3$ nanofluids measured in this paper, the temperatures of the experimental fluids were between 20$^{\circ}$C and 30$^{\circ}$C, the lowest dynamic viscosity of these nanofluids measured under shear flow is 0.3917 Pa$\cdot$S, which is two orders of magnitude higher than the highest dynamic viscosity of 0.001004 Pa$\cdot$S of water at 20$^{\circ}$C$\sim$30$^{\circ}$C. Compared with dynamic viscosity, the volume expansion coefficient, thermal diffusion coefficient, and density of the experimental fluids have extremely small differences compared to water, so these parameters are approximated by the parameter values of water. The temperature difference between the inner and outer cylinders in all measurements in this paper didn't exceed 10$^{\circ}$C. Take the temperature difference $\Delta T$=10 $^{\circ}$C, and substitute it and other parameters into formula (\ref{14}) to calculate the maximum Rayleigh number and get Ra=14.3819 ($ \ll $1000). Therefore, there was no free convection effect in our experiment.
\par In reference \cite{liu2021rheological}, we measured the thermal conductivities of the standard viscosity liquid by using the laboratory-made coaxial cylinder thermal conductivity measuring equipment and compared the measurement results with those of the TC3200L thermal conductivity meter. In this paper, the thermal conductivities of the standard viscosity liquid were measured according to the new theoretical formula (\ref{12}) under the same conditions and the measurement results were compared with those in reference \cite{liu2021rheological}. From Fig. 2, it can be observed that when considering viscous dissipation, the thermal conductivity of the standard viscosity liquid measured and calculated based on both the theoretical formula in reference \cite{liu2021rheological} and the theoretical formula in this paper increases with the increase of shear rate. This is because, from the theoretical formula (\ref{12}), it can be seen that even for Newtonian fluids, the contribution of viscous heat dissipation to thermal conductivity rises with the increase of shear rate. Comparing Figs. 2(a) and 2(b), it is found that the variation of thermal conductivity with shear rate calculated in this paper is weaker than that in reference \cite{liu2021rheological}, and the experimental values of thermal conductivity in this paper are closer to the measured values of TC3200L thermal conductivity instrument. The reason is that, in reference \cite{liu2021rheological}, we regarded the heat passing through the cylindrical side at each radius $r$ position in the annular gap as the heat generated by the inner cylinder heating and the viscous dissipative heat generated throughout the entire annular gap, resulting in an overestimation of the calculated thermal conductivity. In fact, the viscous dissipative heat through the side of the cylinder at radius $r$ only includes the viscous dissipative heat generated from the outer wall of the inner cylinder to the radius $r$. The thermal conductivity calculated according to the theoretical formula (\ref{12}) in this paper is more close to the actual value.
\begin{figure}[!htb]
 \begin{minipage}[t]{0.5\linewidth}
  \centering
  \centerline{\includegraphics[width=6.5cm, height=4.2cm]{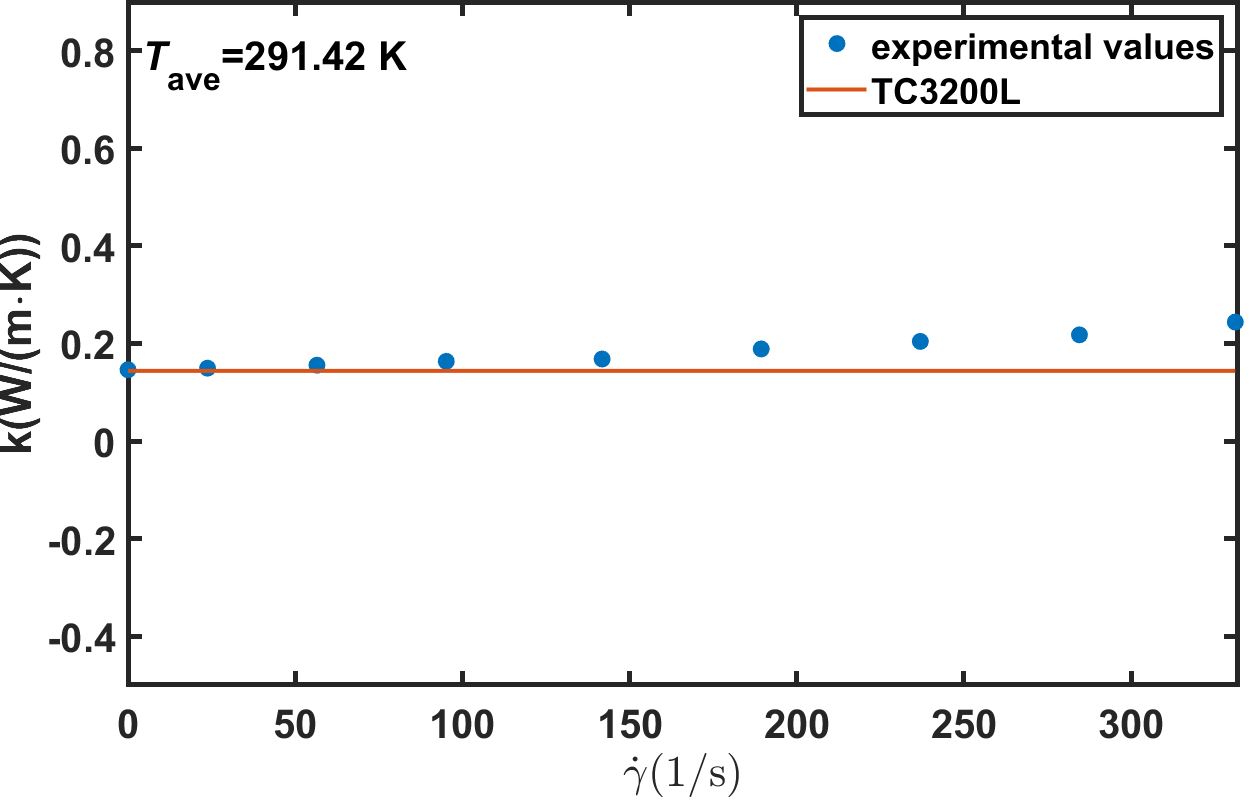}}
 \centerline{(a)}
 \end{minipage}
 \begin{minipage}[t]{0.5\linewidth}
  \centering
  \centerline{\includegraphics[width=6.5cm, height=4.2cm]{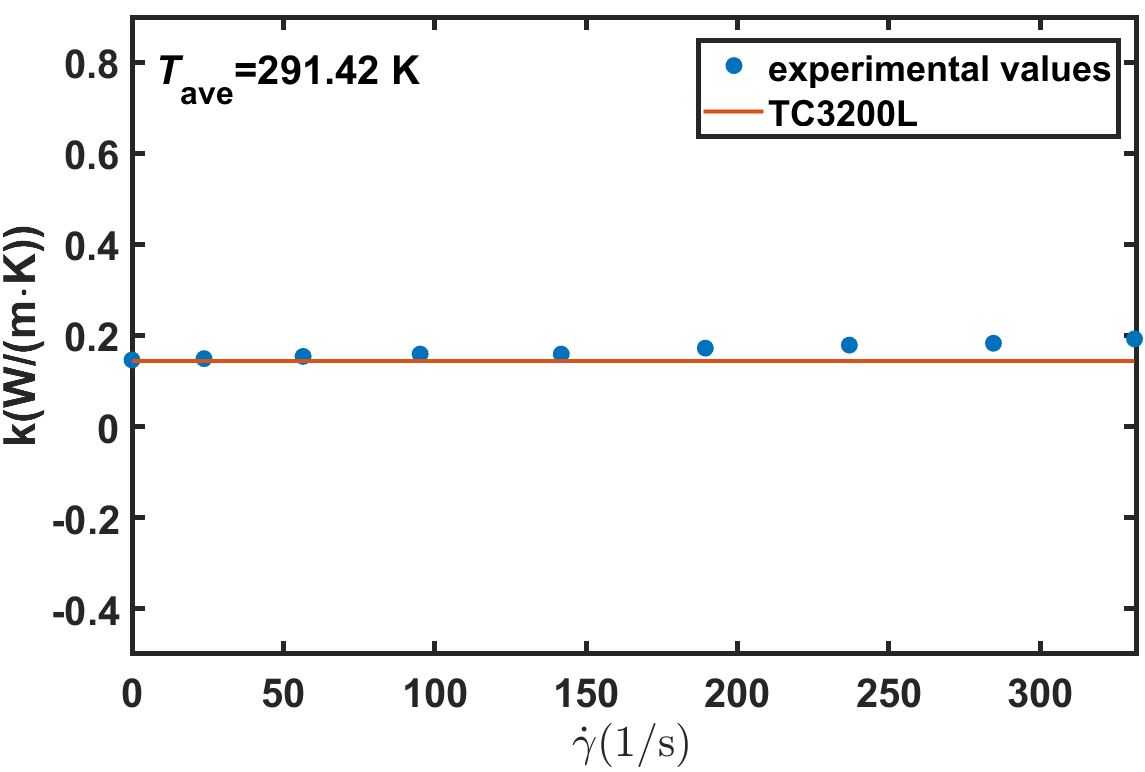}}
  \centerline{(b)}
 \end{minipage}
\caption{Thermal conductivity measurements of standard viscosity liquid in reference \cite{liu2021rheological} (a) and in this paper (b).}
\label{fig14}
\end{figure}
\section{Results and Discussion}
\subsection{{\color{black}Rheological measurements of 2wt$\%$ CMC-based Fe$_{3}$O$_{4}$ and Al$_{2}$O$_{3}$ nanofluids}}
\par Since rheology and thermal conductivity are synergistically correlated, the rheological properties of Fe$_{3}$O$_{4}$ and Al$_{2}$O$_{3}$ nanofluids were measured using the Antonpa MCR 302 rheometer prior to measuring thermal conductivity. Fig. 3 shows the variation of dynamic viscosity with shear rate and volume fraction. From Fig. 3, it can be observed that the shear rate and volume fraction have significant impacts on the dynamic viscosity. The dynamic viscosity decreases with shear rate and increases with volume fraction. Furthermore, the shear change rate of dynamic viscosity reduces with the growth of shear rate and increases with the increment of volume fraction. This means that the rheological non-Newtonian effect decreases with increasing shear rate and rises with growing volume fraction. By comparing the dynamic viscosity of Fe$_{3}$O$_{4}$ nanofluids and Al$_{2}$O$_{3}$ nanofluids, it is found that, the measurement temperature difference between the two nanofluids is only 0.02 K, although the volume fractions of the four groups of Al$_{2}$O$_{3}$ nanofluids are greater than that of Fe$_{3}$O$_{4}$ nanofluids, their dynamic viscosity is still lower than that of Fe$_{3}$O$_{4}$ nanofluids. This indicates that the addition of Fe$_{3}$O$_{4}$ nanoparticles increases the fluid viscosity more significantly than Al$_{2}$O$_{3}$ nanoparticles, which is consistent with the viscosity comparison results between the water-based Fe$_{3}$O$_{4}$ nanofluid measured by Sundar et al. \cite{sundar2013investigation} and the water-based Al$_{2}$O$_{3}$ nanofluid measured by Utomo et al. \cite{utomo2012experimental}.
\par Based on experimental data, an expression (\ref{15}) was proposed for the dynamic viscosity as a function of shear rate and volume fraction. Table \ref{table1} shows the fitting parameters and goodness of fit. From the goodness of fit close to 1, it can be concluded that formula (\ref{15}) with the fitting parameters in Table \ref{table1} can effectively predict experimental values. In addition, the fitting relation (\ref{15}) of dynamic viscosity shows that when the volume fraction is determined, the dynamic viscosity is a function of shear rate and is consistent with the Carreau fluid viscosity model, indicating that the rheological constitutive model of CMC-based Fe$_{3}$O$_{4}$ nanofluids and Al$_{2}$O$_{3}$ nanofluids is the Carreau model. The parameters of the Carreau model are all quadratic polynomials of the nanoparticle volume fraction. When the shear rate is 0, that is, the static state, the dynamic viscosity of the Fe$_{3}$O$_{4}$ and Al$_{2}$O$_{3}$ nanofluids is a quadratic polynomial function of the volume fraction, which is the same form as the viscosity regression equation of the Al$_{2}$O$_{3}$ nanofluid and TiO$_{2}$ nanofluid established by Buongiorno \cite{buongiorno2006convective} using the experimental data of Pak and Cho \cite{pak1998hydrodynamic}. Formula (\ref{15}) also reveals that the variation of dynamic viscosity with volume fraction is influenced by shear rate, which is different from the trend of change at rest.
\begin{equation}
\label{15}
\eta  = \left( {{k_1}{\phi ^2} + {k_2}\phi  + {k_3}} \right){\left( {1 + \left( {{k_4}{\phi ^2} + {k_5}\phi  + {k_6}} \right){{\dot \gamma }^2}} \right)^{\left( {{k_7}{\phi ^2} + {k_8}\phi  + {k_9}} \right)}}
\end{equation}
$k_{1},\ k_{2},\ k_{3},\ k_{4},\ k_{5},\ k_{6},\ k_{7},\ k_{8},\ k_{9}$ are the fitting parameters.
\begin{figure}[!htb]
 \begin{minipage}[t]{0.5\linewidth}
  \centering
  \centerline{\includegraphics[width=5cm, height=3.5cm]{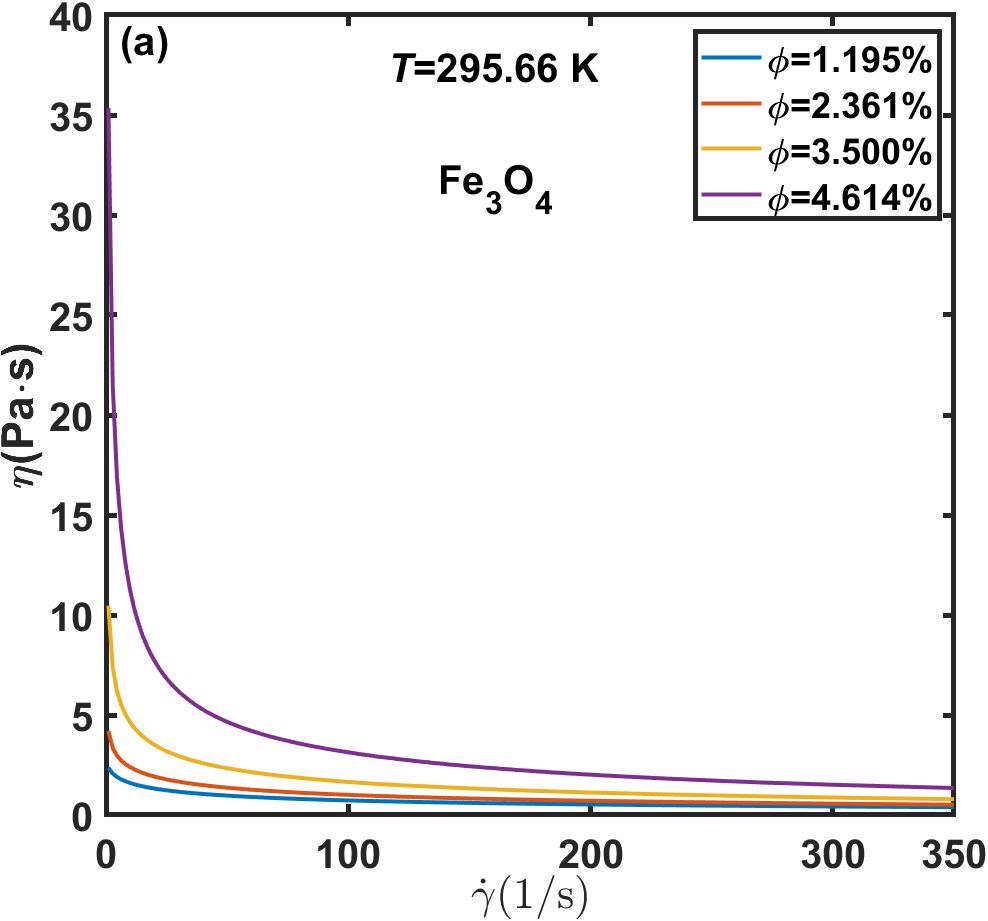}}
 \end{minipage}
 \begin{minipage}[t]{0.5\linewidth}
  \centering
  \centerline{\includegraphics[width=5cm, height=3.5cm]{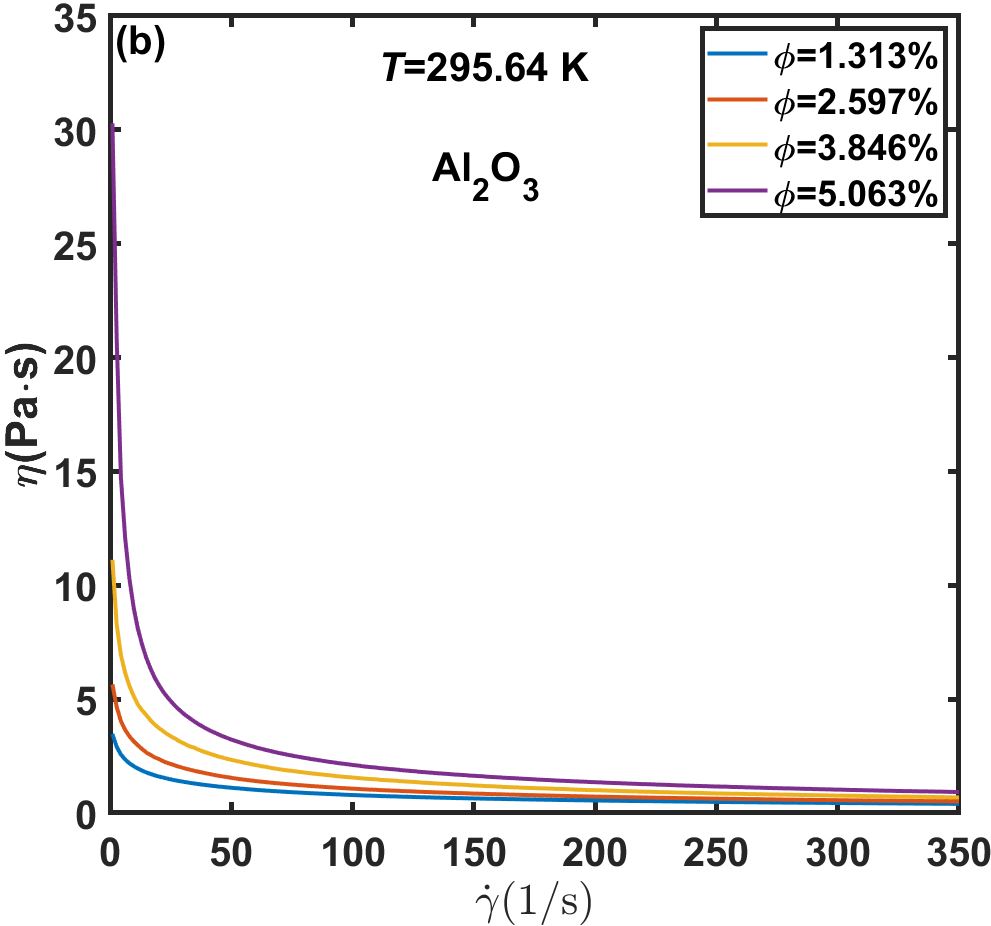}}
 \end{minipage}
\caption{Dynamic viscosity of Fe$_{3}$O$_{4}$ nanofluids (a) and Al$_{2}$O$_{3}$ nanofluids (b).}
\label{fig14}
\end{figure}
\begin{table}[htbp]
	\centering
    \setlength{\abovecaptionskip}{5pt}%
    \setlength{\belowcaptionskip}{10pt}%
	\caption{Fitting parameters and goodness of fit for dynamic viscosity formula (\ref{15}).}
	\label{table1}
    \setlength{\tabcolsep}{2.5mm}{
    \begin{footnotesize}
	\begin{tabular}{llllll}
		\toprule[1pt]
		&$k_{1}$&$k_{2}$&$k_{3}$&$k_{4}$&$k_{5}$\\
		\hline
		Fe$_{3}$O$_{4}$&39307.7990&-1104.1097&13.3430&-26232.7557&116.8388 \\
        Al$_{2}$O$_{3}$&22974.1315&-683.3758&10.8353&-4972.6183&238.2457\\
        \toprule[1pt]
        &$k_{6}$&$k_{7}$&$k_{8}$&$k_{9}$&$R^{2}$\\
        \hline
        Fe$_{3}$O$_{4}$&52.0741&-106.7139&2.7234&-0.1782&0.9969 \\
        Al$_{2}$O$_{3}$&1.3106&-134.4692&5.0583&-0.2358&0.9930\\
        \bottomrule[1pt]
	\end{tabular}
\end{footnotesize}}
\end{table}
\subsection{{\color{black} Analysis of thermal conductivity measurement results}}
For thermal conductivity measurements, the heating power was set at 13 W and the temperature of the circulating water bath was 20$^{\circ}$C. The rotation speed of the outer cylinder was 0$\sim$303 r/min, and the corresponding shear rate was 0$\sim$300.45 1/s. In this paper, the average temperature of the outer wall of the inner cylinder and the inner wall of the outer cylinder is used to approximate the temperature of the experimental fluid. By calculating the average value, the temperature of the Fe$_{3}$O$_{4}$ nanofluids was 295.66 K and Al$_{2}$O$_{3}$ nanofluids was 295.64 K. The temperature settings for rheological measurements of Fe$_{3}$O$_{4}$ nanofluids and Al$_{2}$O$_{3}$ nanofluids in Sec 6.1 were consistent with that of thermal conductivity measurements, respectively.
\par Figs. 4(a) and 4(b) show the experimental values of thermal conductivity of Fe$_{3}$O$_{4}$ nanofluids and Al$_{2}$O$_{3}$ nanofluids, respectively. It is observed from the figures that the thermal conductivity increases with increasing shear rate and volume fraction. And the significance of the change in thermal conductivity with volume fraction/shear rate increases with increasing shear rate/volume fraction. That is to say, there is a mutual promotion effect between shear rate and volume fraction on the enhancement of thermal conductivity. For Fe$_{3}$O$_{4}$ nanofluid, when the volume fraction increases from 1.195\% to 4.614\%, the thermal conductivity rises by 2.6321\% if the shear rate is 0 1/s, and 22.4109\% if the shear rate is 300.45 1/s; In addition, when the shear rate increases from 0 1/s to 300.45 1/s, the thermal conductivity rises by 35.2986\% if the volume fraction is 1.195\%, and 61.3728\% if the volume fraction is 4.614\%. For Al$_{2}$O$_{3}$ nanofluid, when the volume fraction increases from 1.313\% to 5.063\%, the thermal conductivity rises by 5.9798\% if the shear rate is 0 1/s, and 12.8729\% if the shear rate is 300.45 1/s; Moreover, when the shear rate increases from 0 1/s to 300.45 1/s, the thermal conductivity rises by 40.4044\% if the volume fraction is 1.313\%, and 49.5366\% if the volume fraction is 5.063\%.

\par By analyzing the experimental data, the function expression (\ref{16}) of thermal conductivity related to shear rate and volume fraction was presented. Table \ref{table3} depicts the fitting parameters and goodness of fit. The goodness of fit is close to 1 means that expression (\ref{16}) fits well with the experimental data. Figs. 5(a) and 5(b) show the experimental values and fitting curve of Fe$_{3}$O$_{4}$ nanofluids and Al$_{2}$O$_{3}$ nanofluids, respectively. It can also be observed from the figures that the experimental data are evenly dispersed on or around the fitting curves, indicating that the thermal conductivity expression (\ref{16}) with parameters in Table \ref{table3} can well predict the experimental data. The deviation rate between the experimental values and the fitted value is -2.33$\%$$\sim$3.45$\%$ for the Fe$_{3}$O$_{4}$ nanofluids, and -2.15$\%\sim$1.96$\%$ for the Al$_{2}$O$_{3}$ nanofluids. Thermal conductivity expression (\ref{16}) reveals that thermal conductivity is a power law function of shear rate, validating the theoretical conjecture of Pop et al. \cite{subba1993boundary,pop1991mixed}. And the thermal conductivity expression (\ref{16}) can predict both the thermal conductivity in the shear flow state and in the static state. Observing the power-law exponent $m_{3}$ of shear rate in Table \ref{table3}, it is found that $m_{3}$ is less than 1. This indicates that the variation of thermal conductivity with shear rate reduces with increasing shear rate. This is attributed to the fact that, according to the rheological measurements in Fig. 3, the change in dynamic viscosity with shear rate decreases as the shear rate grows, and the non-Newtonian effect of rheology is weakened, which consequently results in the variation of thermal conductivity with shear rate also decreasing, and the non-Newtonian effect of heat transfer is weakened. In addition, the significance of the change in both the dynamic viscosity and thermal conductivity with shear rate is enhanced with increasing volume fraction, implying that the non-Newtonian effects of both rheology and heat transfer increase with increasing volume fraction. There is a consistent synergy between the non-Newtonian effects of rheology and heat transfer, as evidenced by the fact that the non-Newtonian effects of rheology and heat transfer decrease with increasing shear rate and increase with increasing volume fraction.
\begin{equation}
\label{16}
k = {m_1}{\phi ^{{m_2}}}{\dot \gamma ^{{m_3}}} + {m_4}\phi  + {m_5}
\end{equation}
$m_{1},\ m_{2},\ m_{3},\ m_{4},\ m_{5}$ are the fitting parameters.
\begin{figure}[!htb]
 \begin{minipage}[t]{0.5\linewidth}
  \centering
  \centerline{\includegraphics[width=5cm, height=3.5cm]{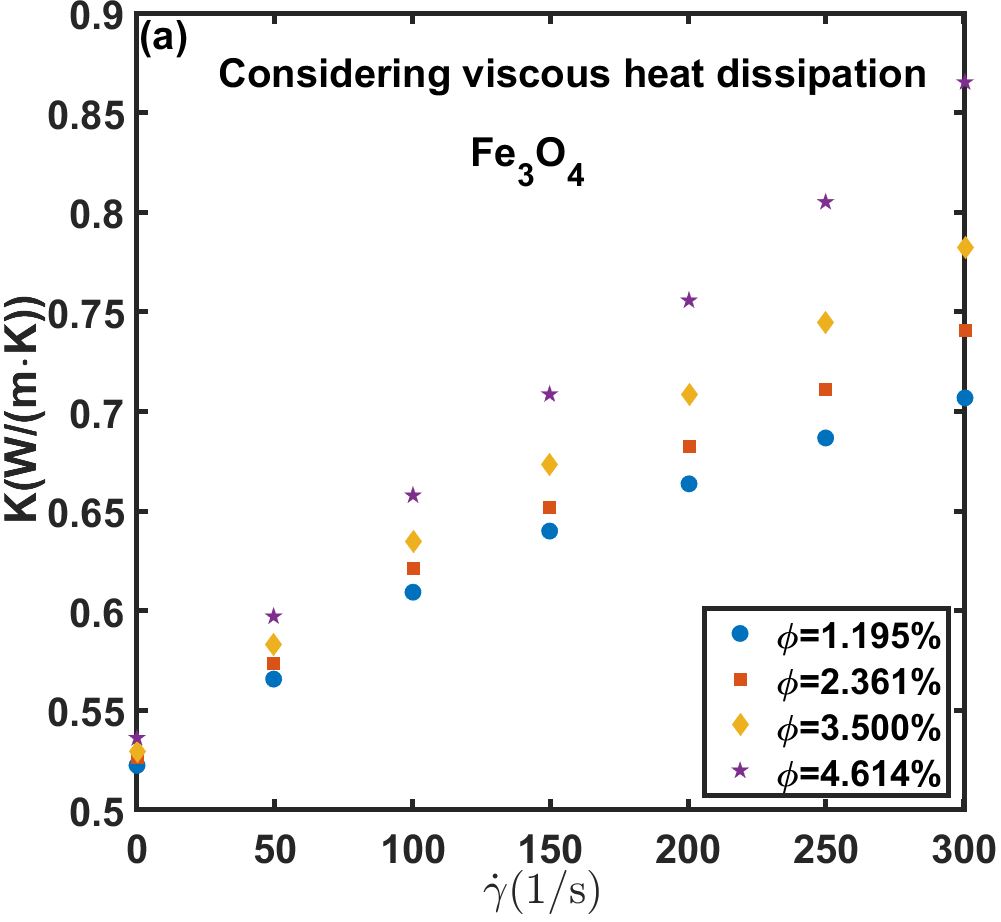}}
 \end{minipage}
 \begin{minipage}[t]{0.5\linewidth}
  \centering
  \centerline{\includegraphics[width=5cm, height=3.5cm]{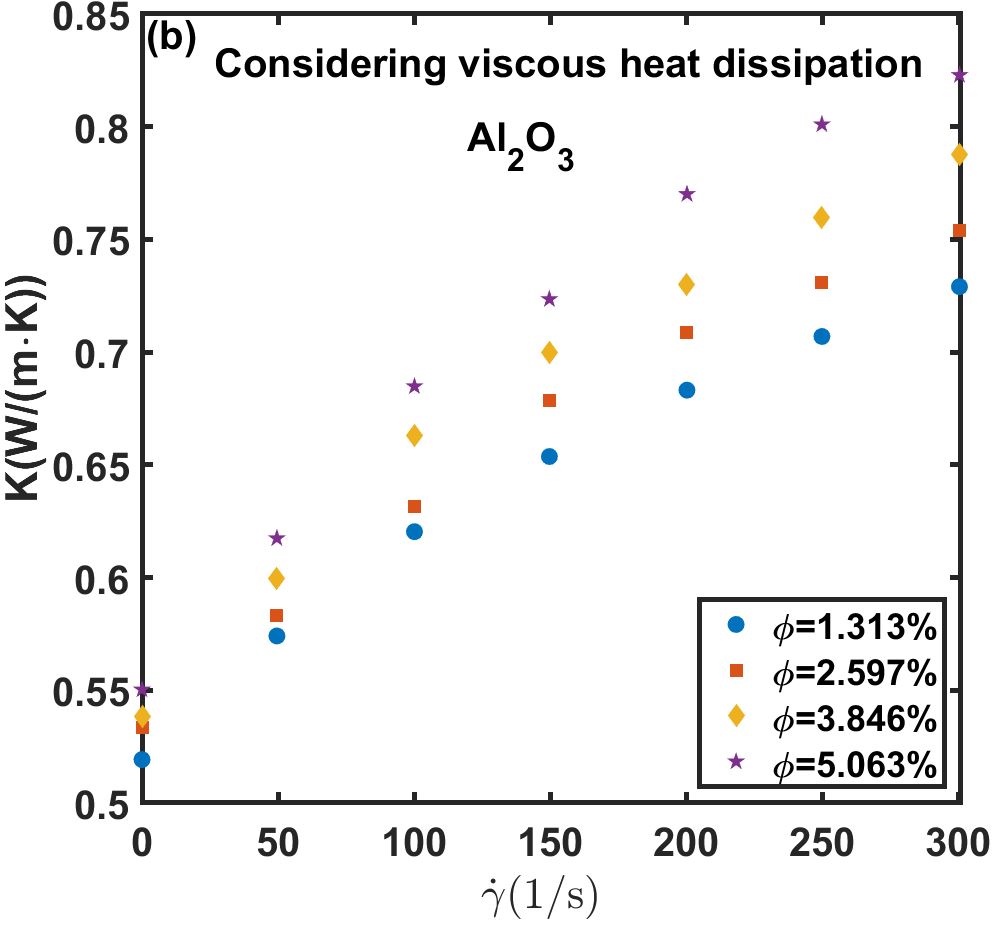}}
 \end{minipage}
\caption{Thermal conductivity of Fe$_{3}$O$_{4}$ nanofluids (a) and Al$_{2}$O$_{3}$ nanofluids (b).}
\label{fig14}
\end{figure}
\begin{figure}[!htb]
 \begin{minipage}[t]{0.5\linewidth}
  \centering
  \centerline{\includegraphics[width=5cm, height=3.5cm]{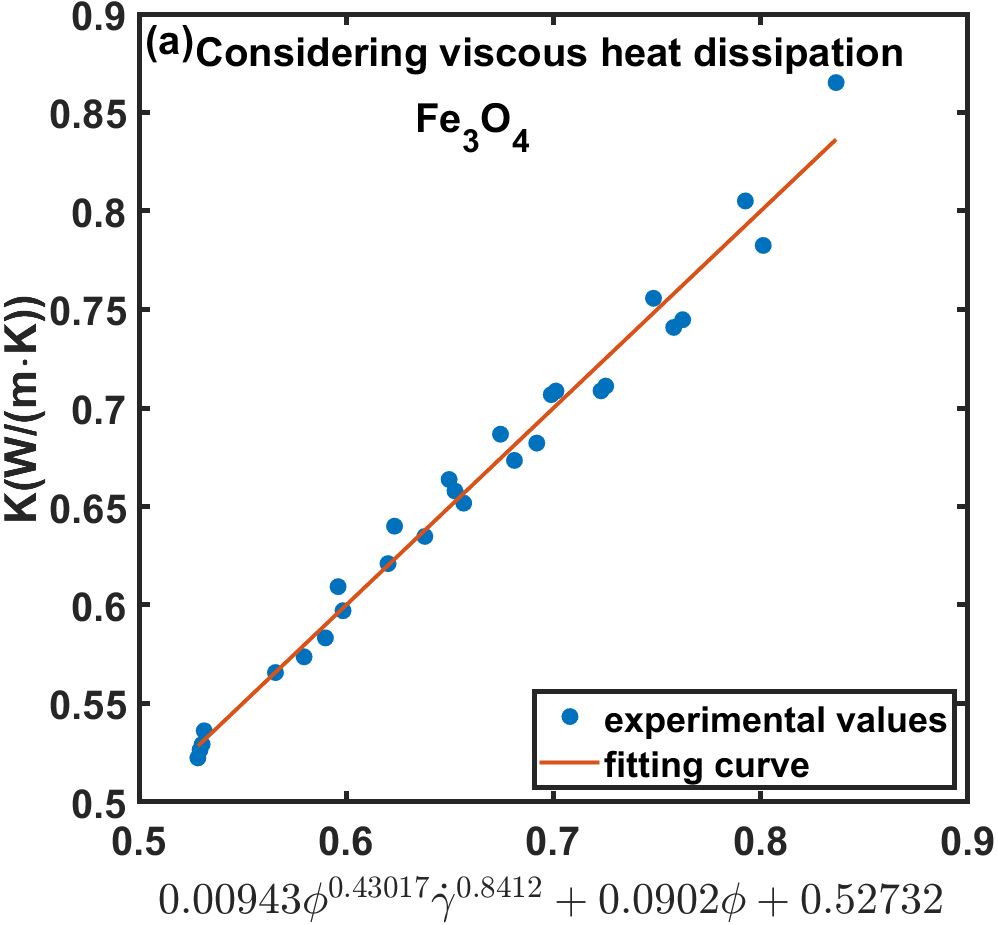}}
 \end{minipage}
 \begin{minipage}[t]{0.5\linewidth}
  \centering
  \centerline{\includegraphics[width=5cm, height=3.5cm]{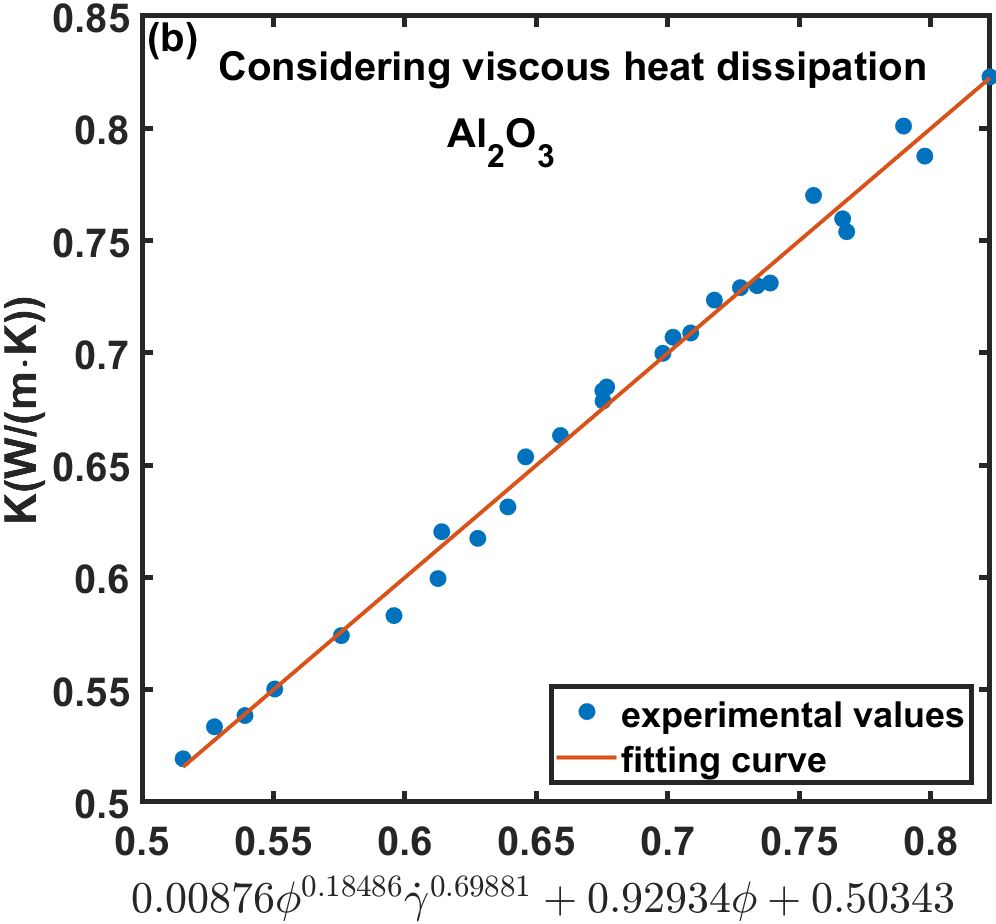}}
 \end{minipage}
\caption{Experimental values and fitted curve for Fe$_{3}$O$_{4}$ nanofluids (a) and Al$_{2}$O$_{3}$ nanofluids (b).}
\label{fig14}
\end{figure}
\begin{table}[htbp]
	\centering
    \setlength{\abovecaptionskip}{5pt}%
    \setlength{\belowcaptionskip}{10pt}%
	\caption{The fitting parameters and goodness of fit.}
	\label{table3}
    \setlength{\tabcolsep}{3mm}{
    \begin{small}
	\begin{tabular}{lllllll}
		\toprule[1pt]
		&$m_{1}$&$m_{2}$&$m_{3}$&$m_{4}$&$m_{5}$&$R^{2}$\\
		\hline
		Fe$_{3}$O$_{4}$&0.00943&0.43017&0.8412&0.0902&0.52732&0.97967 \\
        Al$_{2}$O$_{3}$&0.00876&0.18486&0.69881&0.92934&0.50343&0.99031\\
        \bottomrule[1pt]
	\end{tabular}
\end{small}}
\end{table}
\section{{\color{black}Conclusions}}
In this paper, considering the variation of viscous heat dissipation transmitted radially in the annular gap, we proposed a new theoretical formula for thermal conductivity measurement based on the energy equation of the rotational Couette flow field established in Reference \cite{liu2021rheological}. And four volume fractions of 2wt\% CMC-based Fe$_{3}$O$_{4}$ nanofluids and Al$_{2}$O$_{3}$ nanofluids were prepared. According to the theoretical formula, the rheological properties and thermal conductivity of the prepared nanofluids were experimentally investigated by using the laboratory-made coaxial cylinder measuring equipment \cite{liu2021rheological} and the commercial Anton Paar MCR 302 rheometer. The research results are as follows:\\
(i) Considering the variation of viscous heat dissipation transmitted radially in the annular gap, a new theoretical formula for measuring thermal conductivity considering viscous heat dissipation is derived based on the energy equation of the rotating rotational Couette flow field. The theoretical formula shows that thermal conductivity and rheology have synergistic effect.\\
(ii) The rheological measurement results show that the dynamic viscosity of CMC-based Fe$_{3}$O$_{4}$ nanofluids and Al$_{2}$O$_{3}$ nanofluids decreases with an increase in shear rate and increases with an increase in volume fraction. And the increase of the volume fraction of Fe$_{3}$O$_{4}$ nanofluids is more easily to increase the dynamic viscosity than that of Al$_{2}$O$_{3}$ nanofluids.\\
(iii) The thermal conductivity rises with the increase of shear rate and volume fraction. And the significance of the change in thermal conductivity with volume fraction/shear rate increases with increasing shear rate/volume fraction, which means that there is a mutual promotion effect between shear rate and volume fraction on the enhancement of thermal conductivity.\\
(iv) The non-Newtonian effects of rheology and heat transfer decrease with increasing shear rate and increase with increasing volume fraction, implying that there is a consistent synergy between the non-Newtonian effects of rheology and heat transfer\\
(v) By analyzing experimental data, the expressions of thermal conductivity and dynamic viscosity as functions of shear rate and volume fraction were presented.\\

\begin{tabular}{|ll|}
\hline
\textbf{Nomenclature}&\\
$r$& radius [m]\\
$k$& thermal conductivity [W/(m$\cdot$K)]\\
$T$& temperature of the fluid [K]\\
$u_{\theta}$& tangential velocity [m/s]\\
$\tau_{r\theta}$& shear stress [Pa]\\
$\dot{\gamma}$& shear rate [1/s]\\
$Q_{1}$& heating power of inner cylinder [W]\\
$R_{o}$& inner radius of the outer cylinder [m]\\
$R_{i}$& outer radius of the inner cylinder [m]\\
$h$& effective heating length of the inner cylinder [m]\\
$T_{o}$& temperature of the inner wall of the outer cylinder [K]\\
$T_{i}$& temperature of the outer wall of the inner cylinder [K]\\
$n$& rotational speed of the outer cylinder [r/min]\\
$Q_{2}$& viscous dissipative heat [J/s]\\
$g$& gravitational acceleration [m/s$^2$]\\
$\beta$& volume expansion coefficient [1/K]\\
$\alpha$& thermal diffusion coefficient [m$^2$/s] \\
$\nu$& kinematic viscosity [m$^2$/s]\\
$\eta$& dynamic viscosity [Pa$\cdot$S]\\
$Ra$& Rayleigh number\\
$\phi$& volume fraction\\
\hline
\end{tabular}

{\color{red}\begin{flushleft}
\textbf{}
\end{flushleft}
}

\begin{flushleft}
\textbf{Acknowledgments}
\end{flushleft}

This work is supported by the National Natural Science Foundations of China (Nos. 11772046).
\section*{{\color{black}References}}

\bibliographystyle{ieeetr}
\bibliography{manuscript}

\end{document}